\documentclass[11pt]{iopart}

\usepackage{amsbsy}
\usepackage{iopams}
\usepackage{setstack}
\usepackage{amscd}
\usepackage{amsfonts}
\usepackage{amssymb}
\usepackage{graphicx}
\usepackage{amsgen}

\def\be{\begin{equation}}
\def\ee{\end{equation}}
\def\ba{\begin{eqnarray}}
\def\ea{\end{eqnarray}}
\def\bc{\begin{center}}
\def\ec{\end{center}}

\def\be{\begin{equation}}
\def\ee{\end{equation}}
\def\ba{\begin{eqnarray}}
\def\ea{\end{eqnarray}}
\def\bs{\begin{subequations}}
\def\es{\end{subequations}}
\def\e{ \rm e}

\def\Z#1{_{\lower2pt\hbox{$\scriptstyle#1$}}}
\def\X#1{_{\lower2pt\hbox{$\scriptscriptstyle#1$}}}
\usepackage{color}

\def\ApJ#1{Astrophys.\ J.\ {\bf#1}} 

\begin{document}

\title{Reconstructing a model of quintessential inflation}

\author{Ishwaree P Neupane}

\address{Department of Physics and Astronomy, Rutherford Building,
University of Canterbury, Private Bag 4800, Christchurch 8020, New
Zealand} \ead{ishwaree.neupane@canterbury.ac.nz}

\begin{abstract}
We present an explicit cosmological model where inflation and dark
energy both could arise from the dynamics of the same scalar
field. We present our discussion in the framework where the
inflaton field $\phi$ attains a nearly constant velocity
$m\Z{P}^{-1} |d\phi/dN|\equiv \alpha+\beta \exp(\beta N)$ (where
$N\equiv \ln a$ is the e-folding time) during inflation. We show
that the model with $|\alpha|<0.25$ and $\beta<0$ can easily
satisfy inflationary constraints, including the spectral index of
scalar fluctuations ($n_s=0.96\pm 0.013$), tensor-to-scalar ratio
($r<0.28$) and also the bound imposed on $\Omega\Z{\phi}$ during
the nucleosynthesis epoch ($\Omega_\phi (1 {\rm  MeV})<0.1$). In
our construction, the scalar field potential always scales
proportionally to the square of the Hubble expansion rate. One may
thereby account for the two vastly different energy scales
associated with the Hubble parameters at early and late epochs.
The inflaton energy could also produce an observationally
significant effective dark energy at a late epoch without
violating local gravity tests.

\end{abstract}

\pacs{98.80.Cq, 98.80.-k, 95.36.+x \qquad arXiv:0706.2654}

\maketitle


\section{Introduction}

The WMAP measurements of fine details of the power spectrum of
cosmic microwave background (CMB) anisotropies \cite{WMAP} have
lent a strong support to the idea that the universe underwent an
inflationary expansion in the distant past~\cite{Linde:1990}. The
WMAP data, along with the independent observations of the dimming
of type Ia supernovae in distant galaxies \cite{supernovae} also
favour a result of growing evidence that a large fraction of the
energy density of the present universe is `dark' and has a
negative pressure, thereby leading to the ongoing accelerated
expansion of the universe. It is then natural to ask whether it is
possible to unify the inflation and quintessential fields. In a
viable theory the primordial inflation may lead to have a dark
energy effect in the conditions of concurrent universe. This
picture merits broader discussion.

The main observation that has led many to believe that the dark
energy is Einstein's cosmological constant $\Lambda$, for which
$w_\Lambda\equiv p\Z{\Lambda}/\rho\Z{\Lambda}=-1$ identically and
at all times, is the concordance of different cosmological data
sets, which appear to indicate that the dark energy equation of
state $w\Z{DE} \equiv p\Z{DE}/\rho\Z{DE}$ is not much different
from $-1$ at the present epoch. This solution to dark energy,
however, raises two immediate questions: (i) why is $\rho_\Lambda
\equiv \Lambda/8\pi G \sim 3\rho_M$ today? and (ii) why is
${\rho\Z{\Lambda}}$ ($\sim {10}^{-12}\, {\rm (eV)}^4 $) so tiny?
Apparently, ${\rho^{1/4}\Z{\Lambda}}$ is fifteen orders of
magnitude smaller than the electroweak scale ($m\Z{\rm EW}\sim
{10}^{12}\,eV$), the energy domain of major elementary particles
in standard model physics,  and it is not known at present how to
derive it from other small constants in particle physics.

The cosmological constant as the source of dark energy is only a
possibility. The other possibility is that the cosmological
constant (or gravitational vacuum energy) is fundamentally
variable. Explicit examples are provided by models that use a
dynamical scalar field $\phi$ with a suitably defined scalar
potential $V(\phi)$. Quintessence models are among the most
popular alternatives to Einstein's cosmological constant as they
generally predict at late times a small (but still an appreciable)
deviation from the central prediction of Einstein's cosmological
constant, i.e. $w\Z{\Lambda}=-1$. Observations only require that
$w\Z{\rm DE} < -0.82$ at present epoch~\cite{WMAP,supernovae}, so
one finds worth studying models that support a time-varying dark
energy.

There are arguments in the literature~\cite{Staro00,Nojiri06} that
an appropriate modification of Einstein's theory provides an
alternative resolution to dark energy problem and a natural
framework to address the inflationary paradigm. In this context,
higher-dimensional braneworlds models, scalar-tensor theories and
$R+f(R)$ gravity models, which derive motivations from the
original idea of Kaluza and Klein to its modern manifestation in
string theory, have been of particular interest.

A simple modification of Einstein's theory of general relativity,
which involves a fundamental scalar field $\phi$ with a
self-interacting potential $V(\phi)$, is given by
\begin{equation}\label{E-H}
{\cal L}\Z{E}=  \sqrt{-g}\left( \frac{R}{\kappa^2}-\frac{1}{2}
(\partial\phi)^2-V(\phi)\right)+{\cal L}_m,
\end{equation}
where $\kappa$ is the inverse Planck mass $m_{Pl}^{-1}=(8\pi
G_N)^{1/2}$ and ${\cal L}_{m}$ is the matter Lagrangian. This
theory has been studied over the last three decades by crafting
different types of scalar potentials. The list of the potentials
can be frustratingly long, which includes the quadratic potential
$V(\phi)=\frac{1}{2}\,m_\phi^2 \, \phi^2$ widely considered in
inflationary contexts and the inverse power-law potential $V(\phi)
\propto \phi^{-\alpha}$ ($\alpha\ge 2$). These examples are
perhaps sufficiently simple to understand the basic ideas of
inflation and/or the dynamics dark energy in the concurrent
universe, for a review, see~\cite{Sami:2006}, but they hardly
explain the cosmic expansion of our universe exhibiting all
relevant cosmological properties. It is thus natural to ask
whether it is possible to unify the inflation and quintessential
fields by finding (or constructing) a more general potential.

The model of quintessential inflation~\cite{Peebles:99a} proposed
by Peebles and Vilenkin uses the idea that inflaton potential
could end up as an effective present-day cosmological
constant~\cite{Peebles:88a} or {\it
quintessence}~\cite{Zlatev:1998}. Although quite appealing, the
potential considered in~\cite{Peebles:99a}, which consists of two
parts: $V(\phi)=\lambda (\phi^4+M^4)$ ($\phi<0$) for inflation and
$V(\phi)=\lambda M^8/(\phi^4+M^4)$ ($\phi \ge 0$) for
quintessence, finds no natural field theoretic motivations.
Recently, attempts have been made in constructing a working model
of quintessential inflation within the context of higher
dimensional braneworld models, see,
e.g.~\cite{Sahni:2001,early-quin-in} and references therein for
the earlier proposals. Also, there are suggestions that a
unification of the inflationary era (triggered by $R^2$ type
corrections) and the late-time acceleration can be made through a
simple construction of the modified F(R) models~\cite{Nojiri06},
as well as within the framework of reconstruction of scalar-tensor
gravity~\cite{Staro00}.

In this paper, we reconstruct an explicit observationally viable
model for evolution from inflation to the present epoch by
maintaining the structure of the theory defined by~(\ref{E-H}).
Our reconstruction approach yields a smooth, exponential potential
that describes both the inflation and quintessential parts. The
model can be shown to be compatible with current cosmological
observations, and, presumably, it can be embedded in higher
dimensional theories of gravity, such as string theory.

The rest of the paper is organized as follows. In section 2, we
motivate our construction with an appropriate ansatz for an
inflaton field. We then invert the system of autonomous equations
to determine the inflaton potential, along with other cosmological
variables. There we also find conditions that have to be satisfied
by the reconstructed potential to be consistent with the WMAP
inflationary data. In section 3, we briefly discuss about an
efficient method of reheating, so called `instant preheating',
applicable to our model. In section 4, we include the effect of
ordinary fields and then find an explicit quintessence potential
in a background dominated by radiation (or matter). In section 5,
we show how the reconstructed potential produces an
observationally significant effective dark energy and its
associated late-time cosmic acceleration. In section 6, we discuss
on a possible way of evading local gravity constraints imposed on
the model. Further generalization of our construction with
higher-order corrections is briefly discussed in section 7.
Finally, section 8 is devoted to conclusion.

\section{How might inflaton roll?}

In this section, we neglect the effect of ordinary fields (matter
and radiation). The set of autonomous equations of motion
following from~(\ref{E-H}), with ${\cal L}_m=0$, is given by
\begin{eqnarray}
V(\phi)&=& m\Z{P}^2 H^2 \left[3- {2 m\Z{P}^2 }
\left(\frac{1}{H}\frac{dH}{d\phi}\right)^2 \right],\\
 \frac{\dot{\phi}}{H} &=& 2 m\Z{P}^2 \left(\frac{1}{H}
 \frac{dH}{d\phi}\right),
\end{eqnarray}
where $H\equiv H(\phi)= \dot{a}/a$ is the Hubble expansion
parameter and $a(t)$ is the scale factor of a spatially flat
Friedmann-Robertson-Walker universe.

One of the most crucial parts of a consistent inflationary model
is to understand the time-evolution of the inflaton field $\phi$.
Any choice of $\phi$ should give rise to a flat potential as
required for inflation and also be consistent with cosmological
observations, including WMAP results. To this aim, a simple (and
possibly a natural) choice for the evolution of inflaton field
$\phi$ is
\begin{equation}\label{ansatz-phi1}
\phi \equiv  \phi_0 - \alpha\, m\Z{P} \ln[a/a\Z{i}] - {\gamma}\,
m\Z{P} \left(a/a\Z{i}\right)^{2\zeta},
\end{equation}
where $|\alpha|< {\cal O}(1)$ and $a\Z{i}$ is the initial value of
the scale factor before inflation. We shall take $\gamma=1$ for a
reason to be explained below, while the parameters $\alpha$ and
$\zeta$ ($<0$) will be fixed using bounds on inflationary
variables inferred by the WMAP observations~\cite{WMAP}. The
evolution of the inflaton field in (\ref{ansatz-phi1}), or
equivalently $\phi(t)=\phi_0+ c_0 \ln t + c_1/t^p$ (with $p>0$),
is a generic solution for a modulus and/or dilaton field in many
four-dimensional string models, see, e.g.~\cite{Antoniadis94}. The
assumption (\ref{ansatz-phi1}) holds, almost universally, in many
well motivated inflationary models that satisfy slow roll
conditions, after a few e-folds of expansion. For instance, for
the chaotic model of inflation with the potential $V(\phi)\propto
m^2 \phi^2$, one has $a\propto \e^{\phi^2/2}$ (cf equation~(2.4),
ref.~\cite{Linde:2007fr}) and thus $|\phi| =\sqrt{2}\ln a+ {\rm
const}$. As discussed in~\cite{Wands:06}, even for two scalar
fields model, if the slow-roll conditions $3H \dot{\phi}\Z{i}
\simeq V_{,\,\phi\Z{i}}$ are satisfied at Hubble exit, then ${\cal
N}\equiv \ln a$ depends linearly only on the field values, leading
to a generic situation that $\phi(t) \propto \ln a+$ (small
correction).

The reconstructed scalar field potential is given by
\begin{equation}\label{total-potential}
V(\phi) = m\Z{P}^2 H^2(\phi) \left(3-\epsilon\Z{H}(\phi) \right),
\end{equation}
where
\begin{eqnarray}\label{Hubble-Inf}
H(\phi)&=& M \exp \left[{-\frac{\alpha^2}{2} N(\phi)} -\alpha X
-\frac{\zeta}{2} \,X^2 \right] \nonumber \\
\epsilon\Z{H}(\phi)&\equiv& 2 m\Z{P}^2
\left(\frac{dH/d\phi}{H}\right)^2 = \frac{1}{2} \left(\alpha + 2
\zeta X\right)^2,\label{Hubble-eps}
\end{eqnarray}
where $X\equiv \gamma\,\e^{2\zeta N(\phi)}$, $N(\phi) \equiv \ln
a(\phi(t))+C$. Note that the parameter $\gamma$ appears only in a
combination with $e^{2\zeta N}$; so using a shift symmetry in
$\phi$ and/or choosing the constant $C$ appropriately, we can
always set $\gamma$ to unity, thus $\gamma=1$ henceforth. The
energy scale $M$ (which appears as an integration constant) can be
fixed by the amplitude of density perturbations observed at the
COBE experiments, namely $(dV/d\phi)^{-1} V^{3/2}/(\sqrt{75}\pi
m_{\rm Pl}^3)\simeq 1.92\times 10^{-5}$. With $\alpha\equiv 0.2$
and ${\cal N}_e\equiv \ln (a_f/a_i)\simeq 55$, assuming that
$\zeta<0$, we find $M\simeq 7.4\times 10^{-5} m\Z{P}= 3.1\times
10^{14}~{\rm GeV}$.

With a slowly varying $\epsilon\Z{H}(\phi)$, the scalar curvature
perturbation can be shown to be~\cite{Stewart:1993A}
\begin{equation}
P\Z{\cal
R}^{1/2}(k)=2^{\nu-3/2}\frac{\Gamma(\nu)}{\Gamma(3/2)}(1-\epsilon\Z{H})^{\nu-1/2}
\left(\frac{H^2}{2\pi |\dot{\phi}|}\right)\Z{aH=k},
\end{equation}
where $\nu=3/2+1/(p-1)$ and $a\propto t^p$. The scalar spectral
index $n\Z{s}$ of the cosmological perturbation is defined by
\begin{equation}
n\Z{s}(k)\equiv 1+\frac{d\ln P\Z{\cal R}}{d \ln k}.
\end{equation}
The fluctuation power spectrum is in general a function of wave
number $k$ and evaluated when a given comoving mode crosses
outside the horizon during inflation: $k=a H=a\Z{e} H(\phi)
e^{-\Delta N}$ is, by definition, a scale matching condition.
Instead of specifying the fluctuation amplitude directly as a
function of $k$, it is convenient to specify it as a function of
the number of $e$-folds.

In the case $\alpha= 0$, we get $H(\phi) \propto \exp
\left[\frac{\zeta\kappa^2}{2}\,\phi(2\phi\Z0-\phi)\right]$ and
$\epsilon\Z{H}(\phi)=2\zeta^2 \kappa^2 (\phi-\phi\Z0)^2$. The
scalar potential takes a familiar form: $V(\phi)\propto
m\Z{\phi}^2 \left[3-2\zeta^2\kappa^2 (\phi-\phi\Z0)^2\right]$,
where $m\Z{\phi}^2 \propto H^2$. The number of e-folds is
${\cal N}\Z{\rm e} = \frac{\kappa}{\sqrt{2}}\int_{\phi_2}^{\phi_1}
(\epsilon\Z{H})^{-1/2} d\phi =\frac{1}{2\zeta}\ln
\frac{\phi\Z0-\phi\Z1}{\phi\Z0-\phi\Z2}$,
where $\phi\Z2 < \phi\Z1 <\phi\Z0$. Since $\eta\Z{H}\equiv
\frac{2}{\kappa^2}\, (d^2H/d\phi^2)/H= \epsilon\Z{H}- 2\zeta $ is
small only for a limited range of inflaton values, $\phi\sim
\phi_0$, the number of e-folds is large only when $\zeta$ is very
small. In this case, however, almost no gravitational waves would
be produced, leading to an exponentially suppressed (close to
zero) tensor-to-scalar ratio. The spectrum of scalar (density)
perturbations is also almost Harrison-Zeldovich type, $n_s=1$.
This last result is, however, not consistent with WMAP
observations~\cite{WMAP}. Thus, without loss of generality, we
demand that $|\alpha|>0$; more precisely,
$$ \zeta<0, \quad  -\, 2\zeta e^{2\zeta N }< \alpha < \sqrt{6}
$$ so that $V(\phi)>0$. The spectral index $n_s$ is now given by
\begin{equation}\label{spectral-index}
n\Z{s}-1=2\eta\Z{H}-4\epsilon\Z{H} = -\frac{\alpha^3+6\alpha^2
\lambda +12\alpha \lambda^2 +8 \zeta \lambda + 8
\lambda^3}{\alpha+2 \lambda},
\end{equation}
(up to leading order in slow roll parameters) where $\lambda\equiv
\zeta\e^{2\zeta {\cal N}_e}$ and ${\cal N}_e$ is the number of
e-folds of inflation between the epoch when the horizon scale
modes left the horizon and the end of inflation.

\begin{figure}
\begin{center}
\includegraphics[width=3.6in]{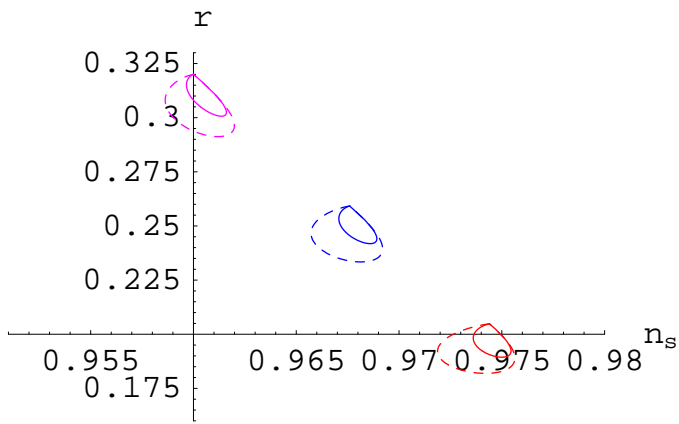}
\caption{\label{ns-vs-r} The tensor-to-scalar ratio $r\simeq 16
\epsilon\Z{H}$ vs the scalar spectral index $n\Z{s}$, with
$\alpha=0.21, 0.20$ and $0.19$ (top to bottom) and
$\zeta=(-0.2,0)$. The solid (dotted) lines are for ${\cal N}_e=60$
(${\cal N}_e=40$).}
\end{center}
\end{figure}
\begin{figure}
\begin{center}
\includegraphics[width=3.6in]{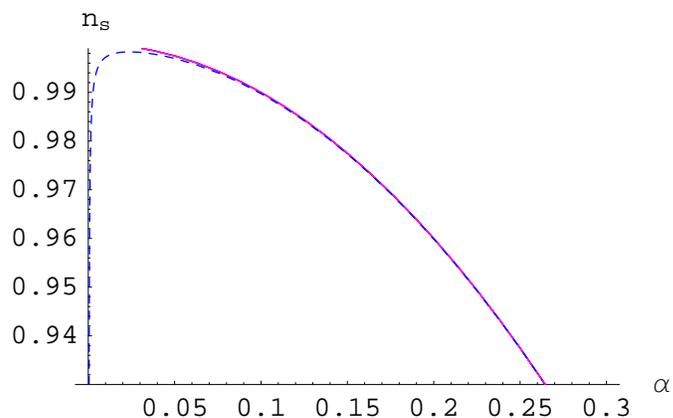}
\caption{\label{ns-vs-alpha} The scalar spectral index $n_s$ vs
$\alpha$, with ${\cal N}_e=70$ (solid line), ${\cal N}_e=40$
(dotted line) and $\zeta=\{-2, 0\}$. Except for $\alpha< |\zeta|
\lesssim 0.05$, the value of $n_s$ does not much depend on
$\zeta$.}
\end{center}
\end{figure}

The scalar spectrum on scales accessible to CMB observations is
perhaps that measured at the instance when observable scales exit
the horizon during inflation. In most models this corresponds to a
phase of inflation between e-folds $50$ and $60$. We summarize the
results in a Table (for ${\cal N}_e=50$ and $\zeta=-0.1$):
\vspace{0.0cm}
\begin{center}
\begin{tabular}{|c|c|c|c|c|}
  \hline
{}   & $n\Z{s} $  & ~$r=16\epsilon\Z{H} $~ & ~$\alpha $~ & ~$\eta\Z{H} $~ \\
  \hline
  $r < 0.28$ & $n\Z{s} \gtrsim 0.965 $
   & $--$ & $ < 0.18 $ & $< 0.017$ \\
  $ n\Z{s}=0.96 $ & $--$ & $0.32$ & $0.200$ & $0.020$\\
  $r = 0.1$ & 0.987 &  $--$  & $0.112$ & $0.006$ \\
  \hline
\end{tabular}
\end{center}

In figure~\ref{ns-vs-r} we show the plot of tensor-to-scalar ratio
$r$ with respect to $n_s$, and in figure~\ref{ns-vs-alpha} the
plot of $n_s$ with respect to $\alpha$. Within our model, both
$n_s$ and $r$ do not much depend on the number of e-folds except
when $\zeta$ is positive, which we reject on physical grounds.

With the WMAP3 bound on the tensor-to-scalar ratio, $r <0.28$, we
find $n\Z{s}\gtrsim 0.965$ for $\zeta\lesssim -0.1$. The bound $r<
0.28$ implies that $\varepsilon\Z{H}< 0.0175$ and imposes a
relation (for a given $N$) between $\lambda$ and $\alpha$. Using
equation~(\ref{spectral-index}), we get $n_s\gtrsim 0.965$ for
$\zeta \le -0.1$. The scalar spectrum is red-tilted except in the
case that $\alpha\lesssim 0$ and both $\zeta$ and $r$ are
sufficiently close to zero, e.g., for $\zeta=- 0.005$ and
$r=0.001$, we get $(\alpha, n_s, N)=(-0.0051, 1.0107, 50),
(-0.0057, 1.0097, 60)$.

\begin{figure}
\begin{center}
\includegraphics[width=3.6in]{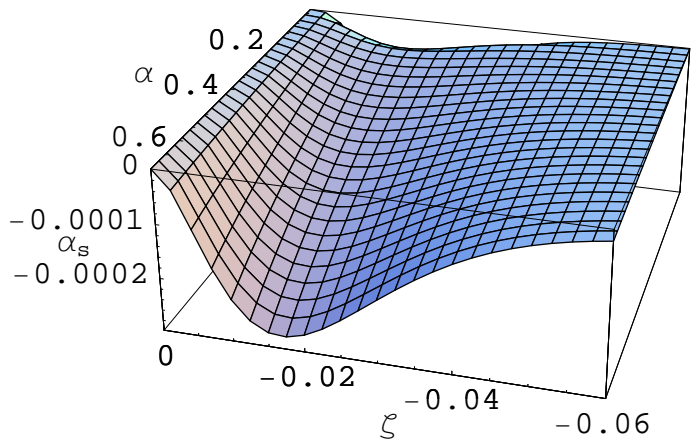}
\caption{\label{3d-running} The running of scalar spectral index,
$\alpha\Z{s}$, with respect to $\alpha$ and $\zeta$ with ${\cal
N}_e=60$. Except for $\alpha< |\zeta| \lesssim 0.05$,
$\alpha\Z{s}$ does not much depend on the number of e-folds.}
\end{center}
\end{figure}

The running of spectral index, $\alpha_s$, is given by
\begin{equation}
\alpha_s \equiv \frac{dn\Z{s}}{d\ln k}=\frac{d n_s}{dN}
\frac{dN}{d\phi}\frac{d\phi}{d\ln k},
\end{equation}
where
\begin{equation}
\frac{d\phi}{d\ln k}=-m\Z{P}
\frac{\sqrt{2\epsilon\Z{H}(\phi)}}{(1-\epsilon\Z{H}(\phi))},\quad
m\Z{P} \frac{dN}{d\phi}= -\frac{1}{\sqrt{2\epsilon\Z{H}(\phi)}}.
\end{equation}
These relations hold independent of our
ansatz~(\ref{ansatz-phi1}). In our model, the value of $\alpha_s$
is found to be small, when satisfying $0.01<\alpha<\sqrt{2}$ and
$\zeta<0$ (cf figure~\ref{3d-running}).

We conclude this section with a couple of remarks. Studies in
\cite{Lyth:1996} show that, in slow-roll inflation, one may relate
the variation of the inflaton in terms of e-folds $N=\ln
(a_f/a_i)$ to the tensor-to-scalar ratio $r$
\begin{equation}
\frac{1}{m_{P}}\frac{d\phi}{dN} = \frac{\phi^\prime}{m_P} =
\sqrt{\frac{r}{8}}
\end{equation}
The WMAP bound on the tensor-to-scalar ratio is $r < 0.28$ ($95\%$
confidence level). This then implies that $\alpha < 0.187$ in the
present construction. This is completely consistent with our
discussion above.

The reconstructed potential may be expressed as
\begin{equation}\label{recons-poten}
V(\varphi)=\frac{H^2(\varphi)}{2\kappa^2}\left[6-\left(\alpha-2\zeta\varphi-2\alpha\zeta
\, N(\varphi)\right)^2\right],
\end{equation}
where
\begin{equation}\label{recons-Hubble}
H(\varphi)\propto \exp\left[ \frac{\alpha^2}{2} N(\varphi)+\alpha
{\varphi}- \frac{\zeta}{2}(\varphi+\alpha N(\varphi))^2 \right],
\end{equation}
where $N(\varphi)=\ln [a(\varphi(t))]$ and $\varphi\equiv
(\phi-\phi\Z{0})/m_P$.
\begin{figure}
\begin{center}
\includegraphics[width=3.6in]{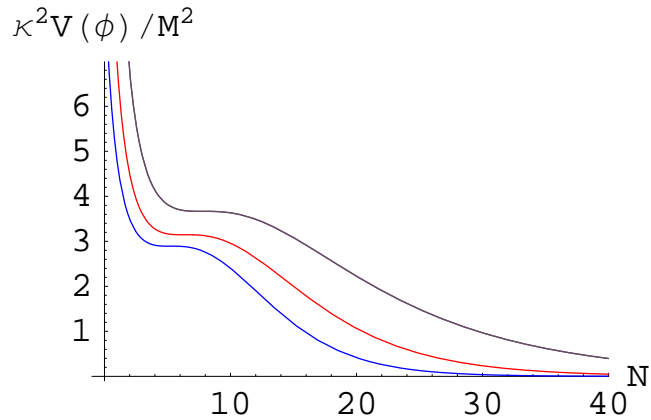}
\caption{\label{full-poten} The scalar potential for some
representative values of $\alpha=0.3,~ 0.4,~ 0.5$ (top to bottom),
$\zeta=-\,0.1$ and $C=- 10$.}
\end{center}
\end{figure}
The shape of the potential (as depicted in
figure~\ref{full-poten}) as well as its functional form is
qualitatively similar to a class of scalar potentials one would
obtain via warped flux compactifications of string theory, see,
e.g.~\cite{Baumann:2007}. The predicted characteristics of
inflationary phase (of the potential) can easily be made to comply
with the WMAP results~\cite{WMAP}. So our method of reconstruction
may be considered as a point in favour of providing a believable
physical basis for the inflation. Moreover, a large part of our
construction does not depend on the details of string theory or
the dynamics of scalar fields abundant in any higher dimensional
theories but has a general validity, and thus would remain useful
even if string theory is invalidated.

\section{Reheating after inflation}

A satisfactory model of inflation should perhaps be followed by a
successful reheating. To this end, the `instant preheating'
mechanism presented in~\cite{Felder:99a} and applied to
exponential potentials in~\cite{Sami04} might perhaps be the most
efficient method for reheating the universe. Here we briefly
outline a viable mechanism of reheating in our model, leaving the
details for future publication.

According to (\ref{ansatz-phi1}), after a few e-folds of
inflation, since $\zeta N<0$, one has $\dot{\phi}\simeq -\alpha\,
m\Z{P} H $. Clearly, with $\alpha<\sqrt{2}$, the kinetic term
never dominates the potential term. As a result there remains the
possibility that the expansion enters inflation from which it
never recovers. So our model has a chance to work only if the
matter and/or radiation energy density terms sometime after
inflation is large enough to dominate the inflaton energy density.

Without loss of generality, we can make the inflation end at the
origin by translating the field
\begin{equation} V(\varphi)= M^2
m\Z{P}^2 (3-\alpha^2/2)\, \e^{\alpha(\varphi/m\Z{P})}+ {\rm small
~correction},
\end{equation}
so after inflation $\varphi\equiv (\phi-\phi_{\rm end})\lesssim
0$. Following~\cite{Felder:99a,Sami04} we assume that the inflaton
field $\varphi$ interacts with another scalar field $\chi$. The
interaction Lagrangian is
\begin{equation}\label{interaction1}
{\cal L}_{\rm int}=-\frac{1}{2} g^2 \varphi^2 \chi^2 - h
\bar{\psi}\psi \chi, \end{equation} where $g$ and $h$ are coupling
constants, and $\psi$ is a Fermi field. The production of $\chi$
particles commences when the adiabatic condition
\begin{equation}
|\dot{m}_\chi|< m_\chi^2
\end{equation}
 is violated, i.e. when
$|\dot{\varphi}|\gtrsim g \varphi^2$, where $m_\chi\equiv
g|\varphi(t)|$. So, the particle production may occur when
\begin{equation}
|\varphi| \lesssim \sqrt{\frac{\dot{\varphi}_{\rm end}}{g}}\sim
\frac{\alpha^{1/2} V^{1/4}_{\rm end}(\varphi)}{3^{1/4}
g^{1/2}}\equiv \varphi\Z{\rm prod}.
\end{equation}
The process of particle production occurs nearly instantaneously,
within the time
\begin{equation}
\Delta t\Z{\rm prod} \sim \frac{|\varphi|}{|\dot{\varphi}\Z{\rm
end}|}\sim \frac{V^{-1/4}_{\rm end}(\varphi)}{\alpha^{1/2}g^{1/2}}
\end{equation}
during which the field $\varphi$ remains in the vicinity of
$\varphi=0$. As the field rolls to $\varphi<0$ direction, the mass
of the $\chi$ particles begins to grow, since $m_\chi\equiv
g|\varphi(t)|$. The occupation number of $\chi$ particles is $
n_k\sim \e^{-\pi (k\Delta t\Z{\rm prod})^2}$, with $k$ being the
canonical momemtum. The energy density of particles of the $\chi$
field created in this process is
\begin{equation}
\rho_\chi=m_\chi n_\chi \left(\frac{a\Z{\rm end}}{a}\right)^3,
\end{equation}
where the number density $n_\chi=(2\pi^3)^{-1} \int_0^\infty k^2
n_k dk \simeq (2\pi)^{-3} (\alpha g)^{3/2} V\Z{\rm
end}^{3/4}(\varphi)$. If the $\chi$ particles can rapidly decay
into fermions or the quanta of the $\chi$ field were to convert
(or thermalize) into radiation, then the radiation energy density
would increase sharply to
\begin{equation} \rho_r \simeq
\rho_\chi \sim \frac{g^{5/2} \alpha^{3/2} V^{3/4}\Z{\rm
end}}{8\pi^3}\, \varphi\Z{\rm prod} \sim 0.0027 g^2\,\alpha^2
V\Z{\rm end}(\varphi).
\end{equation}
At the end of instant preheating
\begin{equation}
\frac{\rho_\varphi}{\rho_r} \sim \frac{370}{\alpha^2 g^2}.
\end{equation}
Although $\rho_\chi/\rho_\varphi$ is small quantity to begin with
(for any generic value of the coupling $g\lesssim 0.3$ and
$\alpha< \sqrt{6}$), $\rho_\chi$ (or the decay product of the
$\chi$ field) will decrease as $a^{-3(1+w)}$ ($w\le 1/3$) and come
to dominate $\rho\Z{\varphi}$ since the field $\varphi$ is rolling
down an exponential potential and its energy density could
decrease much faster $\rho\Z{\varphi}\propto 1/a^6$ after
inflation. To illustrate this one considers a cosmic evolution by
suppressing $dV/d\varphi$, so $\ddot{\varphi} + 3 H
\dot{\varphi}=0$, whose solution is
$\varphi=\varphi\Z{0}+\varphi\Z{1} \int a^{-3} dt$. According to
(\ref{ansatz-phi1}), $\dot{\varphi}\simeq -\alpha\,m\Z{P} H$ and
hence $a(t)=a\Z{\rm end}\left(t\Z{0}+ (3/\alpha) t\right)^{1/3}$.
We then find
\begin{equation}
\dot{\varphi}^2 \simeq \frac{\alpha^2}{3}\,V(\varphi) \sim
10^{-9}\, m\Z{P}^4 \left(\frac{a\Z{\rm end}}{a}\right)^6.
\end{equation}
For $\alpha< \sqrt{3}$, there would be no kinetic regime.
Nevertheless, since $\rho_\chi$ (or the decay products of the
$\chi$ field) may decrease much slower $1/[a(t)]^n$ ($n\le 4$)
than $\rho_\phi$, it will eventually dominate the scalar energy
density before the production of light elements or the BBN epoch.
Instant preheating may be followed by reheating which occurs
through the decay of $\chi$ particles to fermions as is evident
from the interaction term in (\ref{interaction1}).

\section{Growing matter}

Given that the inflaton field $\phi$ decays to some radiation and
heavy particles, it would be natural to expect, at later stages of
inflation, small but nonzero values for both the matter and
radiation energy densities. The growth in matter energy density
can naturally affect (or modify) the form (or shape) of the scalar
potential, leading to an additional term in the potential with a
relatively large slope. This last feature is perhaps required to
make our model compatible with the big-bang nucleosynthesis (BBN)
bound imposed on the scalar field energy density.

Here we take the matter Lagrangian in its simplest form, which is
Einsteinian
\begin{equation}\label{matter-Lag}
{\cal L}_m\equiv {\cal L} (g_{\mu\nu},\psi_m)= \sqrt{-g} \left(
{\rho}\Z{M}+ {\rho}\Z{R}\right),
\end{equation}
where ${\rho}\Z{(i)} \propto a^{-\,3(1+w_{(i)})}$, $i= M$ (matter)
or $R$ (radiation). Of course, one could allow in principle an
explicit coupling between the $\phi$-field and matter. It is
believed that inflation was followed by an instant preheating (or
reheating) and then by a radiation dominated phase, so the
strength of coupling between the field $\phi$ and matter could be
neglected during both the inflationary and the radiation-dominated
epochs. Any such couplings, however, can be relevant at later
stages of evolution, especially, at galactic distance scales (see
section~\ref{Evading}).

The set of autonomous equations of motion that follows from
equations~(\ref{E-H}) and (\ref{matter-Lag}) may be given
by~\cite{Ish06b}
\begin{eqnarray}
\kappa^2 V(\phi) &=&\left[(3+\epsilon)(1-\Omega_w)+\frac{1}{2}\,
\Omega_w^{\,\prime}\right] \,H^2(\phi),
\label{grav-scal1} \\
\kappa^2 {\phi^{\prime}}^2 &=& \Omega_w^{\,\prime}
-2\epsilon (1-\Omega_w), \label{grav-scal2} \\
\Omega_w^{\,\prime} &=& - 2\epsilon\, \Omega_w - 3 (1+w)
\Omega_w,\label{grav-scal3}
\end{eqnarray}
where $\Omega_w\equiv \Omega_M+ \Omega_R$, the prime denotes a
derivative with respect to $N \equiv \ln [a(t)]$, and
\begin{equation}
\phi^\prime =\frac{\dot{\phi}}{H}, ~~ w\equiv \frac{p\Z{R}+
p\Z{M}}{\rho\Z{R}+\rho\Z{M}}, ~~ \epsilon= \frac{H^\prime}{H},~~
\Omega_i \equiv \frac{\kappa^2 \rho_i}{3H^2}.
\end{equation}
During radiation dominance $\Omega\Z{\phi}$ would remain small but
nonzero. This last assumption is consistent with the fact that the
fixed point solution $\Omega_w=1 $ is always unstable. Notice that
the behaviour $V(\phi)\propto H^2(\phi)$ holds also in the
presence of ordinary fields (matter and radiation).

From equations~(\ref{grav-scal1})-(\ref{grav-scal3}), along with
equation~(\ref{ansatz-phi1}), we find
\begin{eqnarray}
\kappa^2 V(\phi)&=& \frac{H^2(\phi)}{\kappa^2}
\left[3(1-\Omega_w)-
B(\phi)\right],\label{recon-tot-po}\\
\epsilon(\phi) &=& -\frac{3}{2} (1+w)\Omega_w -
 B(\phi),\\
\Omega_w &=& \frac{C(\phi)}{C\Z{0}+ 3(1+w)\int (-
C(\phi))\,dN(\phi)},\label{Omega-w}
\end{eqnarray}
where $C\Z{0}$ is an integration constant and
\begin{equation}
B(\phi) \equiv  \frac{1}{2} \left(\alpha+2\zeta \e^{2\zeta
N(\phi)}\right)^2,
\end{equation}
\begin{equation}
 C(\phi)
= \e^{\,\left(\alpha^2-3(1+w)\right)N(\phi)} \exp\left[ 2\alpha\,
\e^{2 \zeta N(\phi)}+\zeta \e^{4\zeta N(\phi)}\right].
\end{equation}
As compared to the inflationary potential given by
equations~(\ref{recons-poten}) and (\ref{recons-Hubble}), we now
have the effect of matter fields (matter and radiation together).
Of course, in the limit that $\Omega\Z{w}\to 0$,
equation~(\ref{recon-tot-po}) reduces to (\ref{recons-poten}).

During radiation domination, since $\Omega\Z{R}\gg \Omega\Z{M}$
and $\Omega\Z{w}\approx \Omega\Z{R}$, we have $w \simeq 1/3$. One
also notes that the last term in equation~(\ref{ansatz-phi1}) does
not contribute (significantly) after inflation. Therefore, from
equations~(\ref{recon-tot-po})-(\ref{Omega-w}), we get
\begin{eqnarray}
\Omega\Z{w} = \frac{1+w-{\alpha}^2/3}{1+ w}\,
\frac{H_0^2}{H^2(\phi)}\, \e^{3(1+w)\kappa
(\phi-\phi_2)/\alpha},\label{Omegaw1}
\end{eqnarray}
where $ H^2(\phi) =  H_0^2 [\e^{\alpha\kappa (\phi-\phi_1)} +
\e^{3(w+1)\kappa (\phi-\phi_2)/\alpha}]$, and
\begin{eqnarray}
V(\phi)= \frac{H_0^2}{\kappa^2} \left[\alpha\Z1
\e^{3(1+w)\kappa(\phi-\phi_2)/\alpha} + \alpha\Z2
\e^{\alpha\kappa(\phi-\phi_1)}\right],\label{double-expo-poten}
\end{eqnarray}
where $\alpha_1 \equiv \frac{\alpha^2}{2}\frac{1-w}{1+w}$,
$\alpha_2 \equiv \frac{6-{\alpha}^2}{2}$, and $H_0$, $\phi_1$,
$\phi_2$ are integration constants; we take $\phi<\phi\Z2\ll
\phi\Z1$. Exponential potentials of a such form, which also arise
ubiquitously in particle physics and string theory
models~\cite{Ish03}, by themselves are a promising ingredient for
building a natural model of quintessential inflation. In order for
the scalar field potential not to dominate the energy density of
the universe during BBN, it is required that $3(1+w)>
\sqrt{6}\alpha$, which easily satisfies the bound imposed on
$\Omega_\phi$ during the nucleosynthesis epoch,
$\Omega\Z{\phi}(1\,{\rm MeV}) \lesssim 0.05$~\cite{Ferreira97a}.
By taking $ \alpha \lesssim 0.8 $ and $w\simeq 1/3$, we correctly
reproduce a double exponential potential anticipated
in~\cite{Barreiro:99a}.

The reason why the quintessential part of the potential,
equation~(\ref{double-expo-poten}), has a different form with
respect to its inflationary part is easy to understand in our
model. During inflation the matter contribution (and its possible
coupling with the inflaton field) can be safely neglected. This
is, however, essentially not the case for quintessence part.
Another source of this difference is that the last term in
equation~(\ref{ansatz-phi1}) does not contribute (significantly)
at later stages of evolution, like during the radiation-dominated
epoch.

\section{Late time acceleration: Dominance of dark energy}

\begin{figure}
\begin{center}
\includegraphics[width=3.6in]{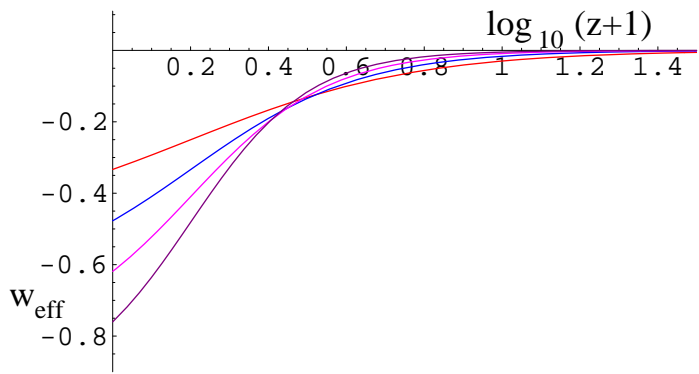}
\caption{\label{Fig2} Evolution of the universe passing from
matter dominance ($w_{\rm eff}\simeq 0$) to scalar field dominance
($w_{\rm eff}<- 1/3$), with $m=3, 5, 10$ and $50$ (from top to
bottom). }
\end{center}
\end{figure}

At late times, without loss of generality, one takes
${\rho}\Z{R}\ll {\rho}\Z{M}$ and $w\simeq 0$. One also assumes
that $\phi$ is rolling only slowly, such that $|\dot{\phi}|/H <
m\Z{P}$. In this case the inflaton potential takes a simpler form,
for the evolution of the universe could naturally lead the
potential part to dominate the kinetic part: $2 V(\phi)\propto
\dot{\phi}^2$, with $m$ being the proportionality constant.
Explicitly, we get
\begin{equation}
V(\phi) = m\Z{P}^2 H(\phi)^2\, \frac{3m}{m+1} (1-\Omega\Z{w}),
\label{inter1}\\
\end{equation}
$\epsilon(\phi) = -\frac{3}{m+1}-\frac{3}{2}\,\widetilde{w}
\Omega\Z{w}$ and $\Omega\Z{w} = 1/[1+ \delta
(z+1)^{-\,3\widetilde{w}}]$, where $z$ is the redshift factor and
$\widetilde{w} \equiv w+ \frac{m-1}{m+1}$. The Hubble parameter
$H(\phi(z))$ (and hence $V(\phi(z))$) can be expressed in a closed
form using the relation $\epsilon=\dot{H}/H^2$. The numerical
constant $\delta$ can also be fixed using observational input: an
ideal situation would be that the universe re-enters into an
accelerating phase ($\epsilon>-1$) for $z\lesssim 1$. The universe
passes from a decelerating phase to an accelerating phase when
$\Omega\Z{w} < \frac{m-2}{2m-1}$. The dark energy equation of
state is $w\Z{\phi}={p\Z{\phi}}/\rho\Z{\phi}=\frac{1-2m}{1+2m}$;
therefore, with $m\equiv 50$, we get $w_{\rm{eff}} \equiv
-1-2\epsilon/3 \sim - 0.76$ and $w\Z{\phi}\sim - 0.98$ (see also
figure~\ref{Fig2}).

\begin{figure}
\begin{center}
\includegraphics[width=3.6in]{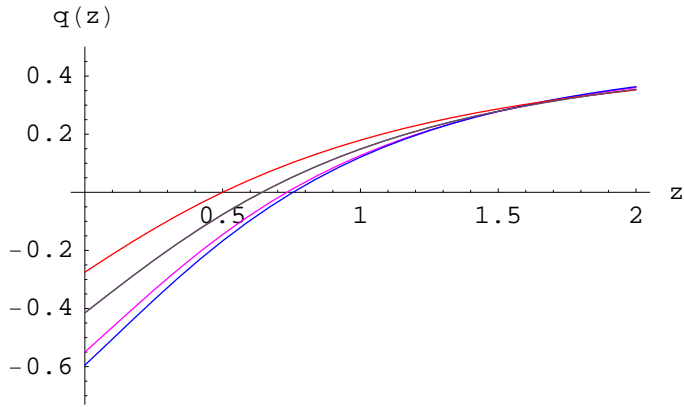}
\caption{\label{Fig3} The deceleration parameter $q(z)$ with
respect to redshift $z$, and $\alpha=0.8, 0.6, 0.3, 0.01$ (top to
bottom). The free parameter $c\Z{0}$ in
equation~(\ref{main-quin-sol}) is chosen such that $\Omega\Z{M
0}\simeq 0.27$.}
\end{center}
\end{figure}

The behaviour of dark energy similar to that depicted in
figure~\ref{Fig2} may be seen directly from
equations~(\ref{Omegaw1})-(\ref{double-expo-poten}). Using the
relation $e^{N}=e^{\ln a}\equiv (1+z)^{-1}$ and making the
assumption that ordinary matter (including cold dark matter) is
approximated by a non-relativistic perfect fluid and $\rho\Z{R}\ll
\rho\Z{M}$, so that $w\approx p\Z{M}/\rho\Z{M} \approx 0$, we find
\begin{equation}
\Omega\Z{w}\simeq \Omega\Z{M}= \left(1-\frac{\alpha^2}{3}\right)
\frac{1}{1+ c\Z{0}\, (1+z)^{\alpha^2-3}}\label{main-quin-sol}
\end{equation}
and
\begin{equation}
\epsilon(z)=-1-q(z) =-\frac{3}{2} \Omega\Z{M}-\frac{\alpha^2}{2},
\end{equation}
where $q$ is the deceleration parameter. Hence
\begin{equation}
H(z) = H\Z{0} \left[ \Omega\Z{m 0} (1+z)^3 + c\Z{0}(1-\Omega\Z{m
0}) (1+z)^{\alpha^2}\right]^{1/2}.
\end{equation}
The numerical coefficient $c\Z{0}$ may be fixed such that
$\Omega\Z{M}=0.27$ at $z=0$. With $\alpha<\sqrt{2}$, the second
term on right-hand side decreases with $z$ at a slower rate as
compared to $\rho\Z{M}$ (which varies as $(1+z)^3$) as well as to
that of the curvature, $\rho\Z{k}$ (which varies as $(1+z)^2$), so
$\Omega\Z{\phi}$ naturally exhibits `dark energy' as late times.
As depicted in figure~\ref{Fig3}, for $\alpha\simeq 0$, the
universe starts to accelerate when $z\lesssim 0.8$. For a larger
$\alpha$, acceleration starts at a lower redshift; with a moderate
value of $\alpha\simeq 0.26$, we get $w\Z{\rm DE}=w\Z{\phi}\simeq
(\alpha^2-3\Omega\Z{\phi})/3\Omega\Z{\phi}\simeq -0.97$.

The present model addresses the cosmic coincidence problem, only
partially. In fact, the cosmic coincidence problem (i.e. why
$\rho\Z{\phi}\simeq 3 \rho\Z{M}$ now?) often involves some kind of
fine tuning, and it is not an exception here. An interesting
observation is that this last phenomenon requires either a
specific ratio between the kinetic and potentials terms, or a
specific value for the field velocity $\phi^\prime\equiv
\kappa\dot{\phi}/H$, which is characterized by the parameter
$\alpha$, so as to realize a quintessence dominance for $z
\lesssim 0.85$.

\section{Evading gravity~constraints}\label{Evading}

In the above discussions we ignored the coupling of the
$\phi$-field with matter. This is perhaps justified.

The dark energy or the cosmic acceleration problem is essentially
a problem associated with largest cosmological scales: in order
for the field $\phi$ to play a role of dark energy its effective
mass should be at least in the range of the present value of the
Hubble parameter, $H\Z{0}\sim 10^{-33} ~{\rm eV}$. In turn, one
takes the runaway quintessence potential satisfying
$\sqrt{V(\phi)}\simeq H\Z{0} \sim 10^{-33}~{\rm eV}$; the range of
the interactions mediated by the scalar field $\phi$ can be of the
order of the Hubble horizon size. However, Newtonian tests of
Einstein's general relativity and fifth force experiments such as
the Cassini satellite experiment put stringent bounds on the
gravitational coupling of light scalar particles. That means, a
putative dark energy field should be sufficiently massive at much
smaller scales. Thus a mechanism similar to that in Chameleon
field theory~\cite{Khoury:03aq}, which combines both a
quintessence-like behaviour leading to dark energy at late time
and a gravitational coupling to matter which is appreciable in
high density regions, could be operative in our model.

To this reason, one allows a nontrivial coupling between the
$\phi$-field and matter, and, accordingly, takes the matter
Lagrangian in a general form
\begin{equation} {\cal L}_m = {\cal L}
(\psi\Z{m}, A^2(\phi) g_{\mu\nu})\equiv \sqrt{-g}\, A^4(\phi)
\sum\widetilde{\rho}\Z{(i)},
\end{equation}
where $\widetilde{\rho}\Z{(i)} \propto
{\hat{a}}^{-3\left(1+w_i\right)}$, $\hat{a}\equiv a A(\phi)$.
$\phi$ couples to the trace of the matter stress tensor,
$g_{(i)}^{\mu\mu} T_{\mu\nu}^{(i)}$, so the radiation term
$\widetilde{\rho}\Z{R}$ does not contribute to the equation of
motion for $\phi$
\begin{equation}
\dot{\rho}\Z{\phi} + 3 H \rho\Z{\phi} \left(1 + w\Z{\phi}\right) =
-\dot{\phi}\eta \alpha\Z{\phi} A(\phi)
\rho\Z{M},\label{nonminimal1}
\end{equation}
where $\eta\equiv (1-3 w_{i})$, $\rho\Z{\phi} \equiv \frac{1}{2}
\dot{\phi}^2 + V(\phi)$, $w\Z{\phi}\equiv p\Z{\phi}/\rho\Z{\phi}$,
$w\Z{i}\equiv p\Z{i}/\rho\Z{i}$ and $ \alpha\Z{\phi} \equiv
\frac{d\ln A(\phi)}{d (\kappa \,\phi)}$. Equation
(\ref{nonminimal1}), along with the equation of motion for
ordinary fluids
\begin{equation}
\dot{\rho}_{i} +3H \rho_{i} (1+ w_{i}) =  + \dot{\phi} \eta
\alpha\Z{\phi} A(\phi) \rho_{i}, \quad (i=M, R),
\label{nonminimal2}
\end{equation}
guarantees the conservation of total energy, namely $\dot{\rho}+
3H (\rho + p)=0$, where $\rho\equiv \rho\Z{M}+\rho\Z{R}+
\rho\Z{\phi}$.

In the discussion below we take $w\Z{M}= 0$. The effective scalar
potential is then given by
\begin{equation}
V\Z{\rm eff}\equiv V(\phi) + {\rho}\Z{M} \int \alpha\Z{\phi}
A(\phi)\, d\phi,
\end{equation}
where ${\rho}\Z{M}\propto 1/a^3$. For $|\alpha\Z{\phi}|> 0$, the
model needs to be confronted with the present-day equivalence
principle bound, $\alpha\Z{\phi}^2\le 5\times 10^{-5}$. On largest
scales probed by WMAP, where ${\rho}\Z{M}\simeq \rho\Z{\rm crit}
\simeq {10}^{-12}~({\rm eV})^4$ (where $\rho\Z{\rm crit} \equiv
3H\Z0^2/8\pi G\Z{N}$ is the critical energy density), the last
term above is only sub-leading, which is suppressed by a factor of
$\alpha\Z{\phi}$. In turn, $\phi$ can be sufficiently light,
$m\Z{\phi}\equiv V\Z{\phi\phi}^{1/2} \sim 10^{-33} \,{\rm eV} \sim
(10^{28} {\rm cm})^{-1}$, and its energy density may evolve slowly
over cosmological time-scales. But within solar system distances,
where ${\rho}\Z{M}$ is roughly $10^{23}$ times larger than its
value on large (Hubble) scales, the term proportional to
${\rho}\Z{M}$ can be more relevant. On Earth, ${\rho}\Z{M}\sim
10^{30}\times \rho\Z{\rm crit}$, the Compton wavelength of the
field $\phi$ can be sufficiently small, $\lambda_c\sim
m_\varphi^{-1} \sim 0.1~{\rm mm} $ as to satisfy local tests of
gravity. That is, in high density (and high curvature) regions the
quintessence field $\phi$ may end up almost in a squeezed state.

\section{Further generalization}

Although the model above is canonical in describing the basic
ideas of quintessence, there exist theoretical and
phenomenological motivations for studying modifications of the
Einstein-Hilbert action which allow non-trivial couplings of
$\phi$ to some quadratic Reimann invariants (of the Gauss-Bonnet
form ${\cal R}^2\equiv
R_{\mu\nu\lambda\sigma}R^{\mu\nu\lambda\sigma}-4 {R_\mu\nu}
R^{\mu\nu} + R^2$) and antisymmetric tensor
fields~\cite{stringy,Ish05d}
\begin{equation}\label{scalar-Lagran2} {\cal L} = \sqrt{-g} \left(
\frac{R}{2\kappa^2} +{\cal L}(\phi)-{\cal F}(\phi) {\cal R}^2-
{\cal G}(\phi) {\cal H}^2 \right)+ {\cal L}_{m},
\end{equation}
where ${\cal L}(\phi) = -\frac{1}{2} (\partial\phi)^2-V(\phi)$,
${\cal H}^2= {\cal H}\Z{\mu\nu\lambda} {\cal H}^{\mu\nu\lambda}$
and ${\cal H}\Z{\mu\nu\lambda}=\partial_{[\mu } B_{\nu\lambda]}$
is the antisymmetric 3-form field strength. Allowing ${\cal
G}(\phi)\ne 0$ in (\ref{scalar-Lagran2}), one introduces a
pseudoscalar degree of freedom $\sigma$, via the ansatz $ {\cal
H}_{\mu\nu\lambda}\equiv \sqrt{g}\,\epsilon_{\mu\nu\lambda\tau}
\partial^\tau \sigma$. Like $\phi$, the axion field $\sigma$
is a function only of time. In particular, the coupling ${\cal
F}(\phi)$ allows new cosmological solutions for which the dark
energy equation of state can be less than $-1$. To be precise, we
note that
\begin{equation} \frac{\kappa^2(\rho\Z{DE}+ p\Z{DE})}{H^2}
= \kappa^2 {\phi^\prime}^2 + (1-\epsilon) \Omega\Z{\cal F}+
\Omega\Z{\cal F}^\prime,
\end{equation}
where $\Omega\Z{\cal F}=8\dot{\cal F} H= 8{\cal F}^\prime H^2$.
The antisymmetric 3-form field does not modify this equation
because it contributes to $\rho\Z{DE}$ and $p\Z{DE}$ with the same
magnitude but with opposite signs, namely,
$\kappa^2\rho\Z{DE}/H^2= x^2/2 + y^2 +3\Omega\Z{\cal
F}+3\Omega\Z{\cal G}$ and $\kappa^2
p\Z{DE}/H^2=x^2/2-y^2-(2+\epsilon)\Omega\Z{\cal F} -\Omega\Z{\cal
F}^\prime-3\Omega\Z{\cal G}$, where $\Omega\Z{\cal G}\equiv 2{\cal
G}(\phi)\,\sigma^{\prime\,2}$, $x\equiv \kappa\dot{\phi}/H$ and
$y\equiv\kappa\sqrt{V}/H$. We can get $w\Z{\phi}\equiv
\rho\Z{\phi}/p\Z{\phi} <-1$, without requiring a superluminal
expansion $\epsilon=\dot{H}/H^2> 0$, or having to introduce a
non-canonical (phantom) field. Most features of the model
(\ref{E-H}) would arise in the limit where ${\cal F}(\phi){\cal
R}^2$ and ${\cal G}(\phi){\cal H}^2$ are sub-leadings to $V(\phi)$
(see below).

In the above model, the axion field $\sigma$ does not play much
role. With a generic choice of ${\cal G}(\Phi)\equiv {\cal G}\Z{0}
e^{2\Phi}$ (where $\Phi\equiv \phi/m\Z{P}$), the $B$-field
equation of motion, $\nabla_\mu\left(e^{2\Phi}
H^{\mu\nu\lambda}\right)=0$, is solved for
$H^{\mu\nu\lambda}=e^{-2\Phi}
\epsilon^{\mu\nu\lambda\tau}\partial_\tau \sigma$. The
integrability condition, $\partial_{[\mu} H_{\nu\lambda\tau]}=0$,
yields $\ddot{\sigma}+3H\dot{\sigma}+2\dot{\Phi}\dot{\sigma}=0$.
With the ansatz~(\ref{ansatz-phi1}), we get
\begin{equation}\label{sol-sigma}
 \frac{m\Z{P}\dot{\sigma}}{H^2}\propto \exp\left[(2\alpha-3)N
+2\zeta e^{2\zeta N} \right].
\end{equation}
After a few e-folds of inflation, the last term above would become
small, since $\zeta N<0$. The scalar potential reads
\begin{equation}
V(\phi)=\frac{H^2}{2} \left[6-\alpha^2 -12 {\cal G}\Z{0}
e^{2(3/\alpha-1)\Phi}\right],
\end{equation}
where $H=H\Z{0}\exp[{\alpha\Phi/2}]$. This result in conjunction
with equations~(\ref{ansatz-phi1}) and (\ref{sol-sigma}) implies
that for $\alpha<2$, ${\cal G}(\phi) {\cal H}^2$ decreases faster
than the scalar potential $V(\phi)$.

Next we briefly discuss some qualitative features of the
reconstructed scalar potential with a nonzero ${\cal F}(\phi)$.
With the ansatz~(\ref{ansatz-phi1}), and with ${\cal G}(\phi)=0$,
the reconstructed potential is given by
equation~(\ref{total-potential}); the parameter
$\varepsilon\Z{H}(\phi)$ reads
\begin{eqnarray}
\varepsilon\Z{H}(\phi) &=& \frac{1}{2}(\alpha+2\zeta
X)^2+3\Omega\Z{\cal F}\nonumber \\
&=&\frac{1}{2}(\alpha+2\zeta X)^2-24 H^2 (\alpha+2\zeta
X)\frac{d{\cal F}(\phi)}{d\phi}.
\end{eqnarray}
Clearly, in the case $\Omega\Z{\cal F}<0$, the coupling ${\cal
F}(\phi)$ could increase the period of inflation by making
$\epsilon\Z{H}$ smaller. This effect can be opposite in the case
$\Omega\Z{\cal F}> 0$: it could be that inflation ended due to a
slowly increasing derivative of the coupling, $d{\cal F}/d\phi$,
such that $\Omega\Z{\cal F}\sim 1/3$.

With ${\cal F}(\phi)\ne 0$, the corresponding potential may be
reconstructed by providing an extra condition or by demanding a
specific relation between the functions $V(\phi)$ and ${\cal
F}(\phi)$ (see, e.g.~\cite{Odintsov06be}, where a general method
of reconstruction was developed, including the effect of
scalar-Gauss-Bonnet coupling). In the particular case that
$a(t)\simeq a\Z{0} e^{H\Z{0} t}$, we find
\begin{equation}
\Omega\Z{\cal F}= -\frac{e^{N(\phi)}}{3H\Z{0}^2}-\alpha^2
-\frac{4\alpha\zeta}{1-2\zeta}\,e^{2\zeta
N(\phi)}-\frac{4\zeta^2}{1-4\zeta}\,e^{4\zeta N(\phi)},
\end{equation}
where $N(\phi)\equiv \ln a(\phi(t))$+ const. Again, after a few
e-folds of inflation, since $\exp[2\zeta N(\phi)]\ll 1$, we get
\begin{equation}
\Omega\Z{\cal F}=-\alpha^2-\frac{e^{N(\phi)}}{3H\Z{0}^2}, \quad
\frac{V(\phi)}{3H^2}=1+\frac{5}{2}\alpha^2 +
\frac{e^{N(\phi)}}{3H\Z{0}^2}.
\end{equation}
This result reveals a generic situation that the coupled
Gauss-Bonnet term is only subleading to the potential $V(\phi)$.
This behavior of our model may be present also when the Hubble
parameter changes appreciably with e-folding time, as happens at
later stages of inflation.

The presence of ordinary fields (matter and radiation) in our
model does not introduce much complication, apart from slightly
modified expressions for $V(\phi)$ and ${\cal F}(\phi)$, for the
added degrees of freedom come with additional equations of motion.

\medskip

\section{Conclusion}

We have presented an explicit cosmological model for evolution
from inflation to the present epoch that we believe satisfies the
main observational constraints, including fine details of the
power spectrum of cosmic microwave background anisotropies, e.g.,
a red-tilted scalar spectrum with small tensor-to-scalar ratio, $r
<0.28$, the bound imposed on $\Omega\Z{\phi}$ during the
nucleosynthesis epoch and present epoch local gravity tests. It is
therefore potentially of great utility.

In our analysis, just one assumption,
equation~(\ref{ansatz-phi1}), that is regarding the evolution of
inflaton field, has been made, which is indeed a common feature of
many motivated slow-roll type inflationary models. Moreover, for a
slowly rolling inflaton field, $m\Z{P} \phi^\prime= m\Z{P}
\frac{d\phi}{d N} < 0.25$, the gravity waves or the amplitude of
tensor perturbations can be suppressed in our model. This might
actually be needed in our model, in order to satisfy the BBN
bound.

The present proposal also simplifies the role of the inflaton by
almost decoupling it from the (background) matter on large
cosmological scales. On the scale of the solar system, due to the
large surrounding matter density, the dark energy field can be
sufficiently massive, e.g., $m\Z\phi \sim \sqrt{\Lambda_{{\rm
eff},\,\phi\phi}} \gtrsim 10^{-3} ~{\rm eV}$, thereby quenching
the deviations from Einstein's gravity on distances larger than a
fraction of millimeter. Moreover, the model possesses an attractor
behaviour for the inflaton and matter densities analogous to the
tracking solution of, e.g., the inverse power-law potential,
$V(\phi)\propto \phi^{-\alpha}$ with $\alpha\ge 2$.

The model proposed here may provide a reasonable explanation to
the question: {\it why is the cosmological vacuum energy small}?
The interpretation of gravitational vacuum energy (or dark energy)
in our framework is {\it likely} to yield $V(\phi) \le
3(1-\Omega\Z{m 0}) H\Z{0}^2 m\Z{P}^2$ and exhibit scaling
behaviour for $\rho_\phi$, being proportional to the square of the
Hubble rate. As a result, within our model, $\rho\Z{\phi} \simeq 2
H\Z{0}^2 m\Z{P}^2 = 2\times 10^{-66}~ {\rm eV}^2 m\Z{P}^2 \simeq
3.5\times 10^{-47}({\rm GeV})^4$ would be the most probable value
of dark energy density at the present epoch.

\bigskip \noindent {\bf Acknowledgments}\\

The author acknowledges the hospitality of the Theory Group at
CERN and DAMTP (University of Cambridge), where part of this work
was carried out. This research is supported in part by the FRST
Research Grant E5229 (New Zealand) and Elizabeth Ellen Dalton
research Award (2007).

\bigskip \noindent {\bf Note added}:
In our model, for $47< N<70$, there also exists a small window in
the parameter space, namely $\alpha =0.011\pm 0.002$ and
$\zeta=-0.03\pm 0.02$, for which $n_s=0.96 \pm 0.01$ and $r\sim
{\cal O}(10^{-3} - 10^{-4}$), see also~\cite{Ish-Scherer} for
other details. This result is compatible with WMAP5
data~\cite{WMAP5}.

\medskip
\section*{References}

\end{document}